\documentclass{emulateapj}
\usepackage{apjfonts}

\usepackage{graphics,graphicx,rotating}
\usepackage{natbib}
\usepackage{xspace}
\usepackage{amssymb}
\usepackage{amsmath}
\usepackage{epstopdf}
\usepackage{subfigure}
\usepackage{color}
\usepackage{soul}

\shorttitle{A Hydrodynamical Model for ``El Gordo''}

\shortauthors{Molnar \& Broadhurst}

\newcommand{\simless} 
     {\ensuremath{\lower 3pt\hbox{$\rlap{\raise5pt\hbox{$\char'074$}}\mathchar"7218$}}}
\newcommand{\simgreat}
     {\ensuremath{\lower 3pt\hbox{$\rlap{\raise5pt\hbox{$\char'076$}}\mathchar"7218$}}}

\newcommand{\simgt}{\lower.5ex\hbox{$\; \buildrel > \over \sim \;$}}
\newcommand{\simlt}{\lower.5ex\hbox{$\; \buildrel < \over \sim \;$}}

\newcommand{\nop}{{\noindent}}

\newcommand{\LCDM}{{\sc $\Lambda$CDM}}

\newcommand{\CHANDRA}{{\sc Chandra}}

\newcommand{\FLASH}{{\sc FLASH}}
\newcommand{\HE}{hydrostatic equilibrium}

\newcommand{\DEGREE}{{$^{\circ}$}}
\newcommand{\ASEC}{\ensuremath{\arcsec}}
\newcommand{\AMIN}{\ensuremath{\arcmin}}

\newcommand{\HMSUN}{{$h^{-1}\,\rm M_\odot$}}

\newcommand{\KMSEC}{{$\rm km\,s^{-1}$}}

\newcommand*{\ltsim}{\ {\raise-.75ex\hbox{$\buildrel<\over\sim$}}\ }
\newcommand*{\gtsim}{\ {\raise-.75ex\hbox{$\buildrel>\over\sim$}}\ }
\newcommand*{\proptosim}{\ {\raise-.75ex\hbox{$\buildrel\propto\over\sim$}}\ }

\newcommand{\CLJMAS}{{\sc CL J0152-1347}}
\newcommand{\MACSJSEVEN}{{\sc MACS J0744.8+3927}}
\newcommand{\CLJONETWO}{{\sc CL J1226.9+3332}}
\newcommand{\MACSJOSEVEN}{{\sc MACS J0717.5+3745}}

\begin{document}

\title{A Hydrodynamical Solution for the ``Twin-Tailed'' Colliding Galaxy Cluster ``El Gordo''}

\author{
Sandor. M. Molnar\altaffilmark{1} \& Tom Broadhurst\altaffilmark{2,3}
}

\altaffiltext{1}{Department of Physics, 
                      National Taiwan University, Taipei 10617, Taiwan, sandor@phys.ntu.edu.tw}

\altaffiltext{2}{Department of Theoretical Physics, University of the Basque Country, Bilbao 48080, Spain}
                                            
\altaffiltext{3}{Ikerbasque, Basque Foundation for Science, Alameda Urquijo, 36-5 Plaza Bizkaia 48011, Bilbao, Spain}

\begin{abstract}
The  distinctive cometary X-ray morphology of the recently discovered massive galaxy cluster ``El Gordo'' (ACT-CT J0102-4915; $z$~=~0.87) indicates that an unusually high-speed collision is ongoing between two massive galaxy clusters. A  bright X-ray ``bullet'' leads a ``twin-tailed'' wake, with the SZ centroid at the end of the Northern tail. We show how the physical properties of this system can be determined using our \FLASH-based,  N-body/hydrodynamic model, constrained by detailed X-ray, Sunyaev--Zel'dovich (SZ), and Hubble lensing and dynamical data. The X-ray morphology and the location of the two Dark Matter components and the SZ peak are accurately described by a simple binary collision viewed about 480 million years after the first core passage. We derive an impact parameter of $\simeq\,$300 kpc, and a relative initial infall velocity of $\simeq\,$2250~\KMSEC\ when separated by the sum of the two virial radii assuming an initial total mass of $2.15 \times 10^{15}$~${\rm M}_{\odot}$ and a mass ratio of 1.9. Our model demonstrates that tidally stretched gas accounts for the Northern X-ray tail along the collision axis between the mass peaks, and that the Southern tail lies off axis, comprising compressed and shock heated gas generated as the less massive component plunges through the main cluster. The challenge for \LCDM\ will be to find out if this physically extreme event can be plausibly accommodated when combined with the similarly massive, high infall velocity case of the Bullet cluster and other such cases being uncovered in new SZ based surveys.
\end{abstract}


\section{Introduction}
\label{S:Intro}

Collisions between galaxy clusters are the most energetic events in the cosmos, with unique implications for structure-formation and the nature of dark matter.
The concordance \LCDM\ model, predicts that the infall velocities of massive merging systems are typically less than 1000 \KMSEC\ \citep{ThomNaga2012MNRAS419}.
However, a few massive merging clusters discovered recently suggest a significant tail in the velocity distribution at high infall velocities (\simgreat$\;$3000 \KMSEC). 
The Bullet cluster (1E0657-56) is readily identified as an extreme example by its clearcut Mach cone found in its X-ray image \citep{MarkevitchET02}. 
Velocities derived from X-ray observations using the shock jump conditions and N-body/hydrodynamical 
simulations support this interpretation \citep{MastBurk08MNRAS389p967,SpringelFarrar2007MNRAS380,MarkevitchET2004ApJ606}, but with inferred velocities in the range 2700--4500 \KMSEC. 
Several other merging clusters have been subsequently found with a bullet-cluster-like morphology (A2744: \citealt{OwersET2012ApJ750L}; A2146: \citealt{RusselET201004.1559}), and some also identified as high-infall-velocity systems
(\CLJMAS: using detailed hydrodynamical simulations; \citealt{Molnet2012ApJ748}; 
\MACSJOSEVEN: using radial velocities and the kinematic SZ effect; \citealt{SayersET2013ApJ778,Mroczkowski2012APJ}).
These high-infall-velocity merging clusters provide a potential challenge for the concordance \LCDM\ model. 
However, to test the concordance model with these mergers, we need more accurate and robust determination of impact velocities based on multifrequency observations and detailed numerical simulations, as well as larger volume cosmological simulations to define the probability of these extreme merging systems
\citep{WatsonET2014MNRAS437,ThomNaga2012MNRAS419}.

The newly discovered "El gordo" galaxy cluster (ACT-CL J0102-4915, $z=0.87$) by the Atacama Cosmology Telescope (ACT) Sunyaev--Zel'dovich (SZ) cluster survey (Marriage et al. 2011) has the largest SZ decrement of any cluster reported  and displays a bullet-like X-ray morphology in deep follow-up Chandra images \citep{MenanteauET2012ApJ748}.
Multifrequency observations of this cluster show a large offset between the X-ray and SZ centroids, and between the X-ray and the centroid of mass-surface density of the main cluster component \citep{ZitrinET2013arXiv1304.0455,JeeET2013arXiv1309.5097}.
These large offsets are simple distinguishing features caused by high velocity encounters as shown by \cite{Molnet2012ApJ748} using N-body/hydrodynamical simulations, suggesting that El Gordo is a high-infall-velocity merging cluster.
The lensing work  identified El Gordo as the most massive cluster known at z$\;$\simgreat$\;$0.6, as of today \citep{JeeET2013arXiv1309.5097,MenanteauET2012ApJ748}.

In this paper we present results from numerical simulations of idealized binary encounters between galaxy clusters containing dark matter and gas initially in equilibrium based on the adaptive mesh code, \FLASH.
Our main goal is to find a physical interpretation of the overall features of El Gordo found in multifrequency observations. 
The structure of this paper is the following. 
In Section~\ref{S:ELGORDO} we summarize results from previous analyses of El Gordo based on multifrequency observations and numerical simulations. Then, in Section~\ref{S:FLASH}, we describe our \FLASH\ simulations, and methods to obtain simulated X--ray and SZ images, and mass surface density maps. We present our results and provide a physical interpretation of the morphology of multifrequency observations of El Gordo in Section~\ref{S:Results}. 
We discuss our results in comparison with previous investigations in Section~\ref{S:Discussion}. 
We summarize our findings and their implications to cosmology in Section~\ref{S:Conclusion}.

Unless stated otherwise, errors quoted are 68\% CL.
We adopt, as our concordance cosmological model, the 5-year WMAP cosmology, $\Omega_m = 0.23$, $\Omega_\Lambda = 0.73$, and H$_0~=~70.50$ km s$^{-1}$ Mpc$^{-1}$ \citep{KomatsuET2009ApJS180}. This model gives the scaling of 1\AMIN\ corresponds to 459.5 kpc
(note, we do not use the dimensionless Hubble parameter, $h$, because numerical simulations work with physical distances).

\section{Multi-frequency observations of El Gordo}
\label{S:ELGORDO}

El Gordo has the largest SZ decrement in the 455 deg$^2$ ACT SZ cluster survey (Marriage et al. 2011), and the most significant detection in the overlapping 2500 deg$^2$ South Pole Telescope (SPT) Sunyaev-survey (Williamson et al. 2011). 
It was confirmed as a high redshift ($z$ = 0.870) galaxy cluster by optical observations \citep{MenanteauET2012ApJ748,MenanteauET10}.
The amplitude of the SZ decrement of El Gordo is comparable to the massive Bullet cluster, and it was found to be the most massive, X-ray, and SZ right cluster at z$\;$\simgreat$\;$0.6 discovered as of today \citep{MenanteauET2012ApJ748}.

A detailed study of this cluster using multifrequency observations was presented by 
\cite{MenanteauET2012ApJ748}.
Chandra ACIS-I observations of El Gordo reveal a disturbed morphology.
The X-ray emission of the cluster is elongated in the South--East to North--West direction. 
One of the tails is following the peak emission at an angle of about 45\DEGREE\ towards North-West (Northern tail), 
below that is located the Southern tail, which may be interpreted as a wake from a merging event. 
Menanteau et al. found a high overall X-ray temperature of T$_X$ = $14.5 \pm 0.1$ keV, and a peak temperature of $\sim 20$ keV around the shock region.
Based on the X-ray morphology, the high X-ray temperature and SZ amplitude, Menanteau et al. suggested that El Gordo is a high-infall-velocity binary merger at high redshift, very similar to the Bullet cluster located at a lower redshift.

The relative positions of the mass surface density, the X-ray and SZ peaks are important in the physical interpretation of a merging galaxy cluster such as El Gordo. \cite{MenanteauET2012ApJ748} located two main peaks in the images of the galaxy number density, the rest frame $i$-band and 3.6~$\mu$m luminosity density, and stellar mass density (derived from infrared IRAC/Spitzer imaging using a spectral energy distribution fitting procedure), which are correlated with the mass distribution.
They concluded that the NW mass component centered on the Northern tail is more massive than the mass center in the SE, in the vicinity of the X-ray peak. The SZ peak was found to be close to the NW component slightly offset toward the SE component.
Their results suggest a mass ratio of about 2:1 of the two components.
Based on the cometary morphology, the high X-ray temperature, the mass distribution, and the offsets between mass peaks and the X-ray and SZ peaks, Menanteau et al. suggest that the bright 
X-ray peak marks the emission from an infalling cluster moving from the NW towards SE plunging through the main cluster disrupted in its X-ray emission indicating a large infall velocity producing a wake with two tails.

Using mass scaling relations based on X-ray, SZ observations and velocity dispersion, Menanteau et al. (2012) estimated a total virial mass of El Gordo to be $1.51\pm 0.22 \times 10^{15}$ \HMSUN. They found that all of their mass estimates are less than 2$\sigma$ away from the interval of 1.2--2.0$\times 10^{15}$ \HMSUN. 
This mass estimate for El Gordo is compatible with subsequent analyses of strong and weak lensing data: 
$1.6 \times 10^{15}$ \HMSUN\ \citep{ZitrinET2013arXiv1304.0455}, and 
$2.15 \times 10^{15}$ \HMSUN\ \citep{JeeET2013arXiv1309.5097}, respectively.

Based on the X-ray morphology and that the location of the peak of the galaxy number distribution of the SE component 
which precedes the X-ray peak by 173 $h^{-1}$ kpc, 
Menanteau et al. suggest that El Gordo is a high redhsift close analogue to the bullet cluster.
In this scenario, as a consequence of the high infall velocity, 
the gas (marked by the X-ray peak) trails the dark matter associated with the infalling cluster (marked by the peak of the galaxy number distribution) due to ram pressure.
However, this close analogy with the bullet cluster is questionable, because another mass proxy, the peak of the $i$-band luminosity and the mass peak of the infalling cluster from weak lensing \citep{JeeET2013arXiv1309.5097} are, in fact, trailing the X-ray peak. 
The positions of the mass peaks estimated using different methods scatter around the X-ray peak within a 173 kpc, thus we conclude that the offset between the X-ray peak and the center of the infalling cluster has not been established.

Recently discovered massive high redshift clusters and large impact velocity merging clusters pose a potential challenge to the concordance \LCDM\ model \citep{ThomNaga2012MNRAS419,LeeKomatsu2010ApJL718}. 
\cite{MenanteauET2012ApJ748} and \cite{JeeET2013arXiv1309.5097} estimated the probability to find a cluster with the mass and the  redshift 
of El Gordo based on cosmological numerical simulations using the exclusion curve method of \cite{MortonsonET2011PRD83}, 
and concluded that, although a cluster such as El Gordo is a rare massive cluster, it is within the allowed mass range predicted by the \LCDM\ models.
However, \cite{JeeET2013arXiv1309.5097} noted that a more accurate mass measurement may put El Gordo into the excluded region.

The discovery of the Bullet cluster with its high inferred infall velocity provided the first challenge to the concordance \LCDM\ model 
\citep{MastBurk08MNRAS389p967}.
Estimates based on cosmological simulations found that the probability of finding a cluster with the infall velocity implied by the bullet cluster is very unlikely \citep{ThomNaga2012MNRAS419,LeeKomatsu2010ApJL718,HayashiWhite2006MNRAS.370L}.
\cite{ThomNaga2012MNRAS419} used cosmological simulations with different box sizes (0.25--2 h$^{-1}$ Gpc) to improve the accuracy of the probability distribution at the high end of the velocity tail taking into account the effect of the box size. 
They confirmed that it is very unlikely ($3 \times 10^{-8}$) to find even one massive cluster merger with impact velocity of \simgreat$\;$3000 \KMSEC\ in a concordance \LCDM\ model at the redshift of the Bullet cluster. 
The most probable infall velocity was found to be about 550 \KMSEC, (see Figure 15 of \citealt{ThomNaga2012MNRAS419}). 
Most recently \cite{WatsonET2014MNRAS437} carried out a simulation using large box size (6 h$^{-1}$ Gpc) to improve on the statistics of large clusters and rare cosmological events. 
They found no merging system with the same large relative velocity, spatial separation, and redshift as the Bullet cluster (see their Figure 7). 
They conclude, however, that this might be due to their low statistics of these rare major mergers at a fixed redshift.
Clearly, more numerical work is necessary to quantify the probability distribution of relative velocities of mergers as a function of redshift, masses, and distances between the components.

Since the discovery of the Bullet cluster several other merging clusters have been found with large offsets between the X-ray peak and 
the SZ and dark matter surface density peaks (MACS J0025.4-1222: \citealt{BradacET2008ApJ687};
\CLJMAS: \citealt{Massardi2010APJL718}; 
A2163: \citealt{BourdinET2011AA527}; 
\MACSJSEVEN\ and \CLJONETWO: \citealt{Korngut2011ApJ734};
\MACSJOSEVEN: \citealt{Mroczkowski2012APJ};
DLSCL J0916.2+2951: \citealt{Dawsonet2012ApJl747};
SL2S J08544-0121: \citealt{GastaldelloET2014arXiv1404.5633}).
These offsets are generated by the different behavior of the collisionless component, the dark matter, and the gas during the merging process:
the dark matter responds only to gravity, but the gas is subject to hydrodynamical effects, and that the X-ray emission and the SZ effect depend differently on the gas density and temperature. 

The size of the offsets between the dark matter and the gas peaks is determined by the relative strength of the gravitational force which is trying to keep the gas locked into the potential well of the dark matter, and the ram pressure which is removing gas from the infalling subcluster.
Using N/body--hydrodynamical simulations of merging galaxy clusters, \cite{Molnet2012ApJ748} showed that a large impact velocity is necessary to explain hundreds of kpc scale offsets.

The relative velocity of El Gordo was estimated by \cite{MenanteauET2012ApJ748} from the measured line-of-sight peculiar velocity difference between the SE and NW galaxy concentrations, as 586$\pm$96 \KMSEC.
However, for the BCG, they found a higher velocity difference: 731$\pm$66 \KMSEC.
Assuming a projection angle relative to the plane of the sky of $\theta$, the relative velocity can be expressed as 586/$\sin \theta$. 
Based on the X-ray morphology, Menanteau et al. argue that $\theta$ should be small, and they obtain a relative velocity of 2300 \KMSEC\ and 1200 \KMSEC\ assuming 15\DEGREE\ and 30\DEGREE\ for $\theta$, and use this to estimate the infall velocity. 
The location of radio relics ahead and behind the two mass centers enclosed by the detected X-ray emission support this interpretation \citep{LindnerET2014ApJ786}. 
These estimated infall velocity values fall on the less likely to the unlikely region of the velocity probability distribution derived by 
\cite{ThomNaga2012MNRAS419}. Menanteau et al. conclude that a more accurate determination of the infall velocity is necessary for El Gordo using numerical simulations to find out wether it is allowed by the concordance \LCDM\ model.

Binary merger simulations to reproduce the observed properties of El Gordo using 
a publicly available smoothed particle hydrodynamics (SPH) code GADGET-3 
(Springel 2005; Dolag \& Stasyszyn 2009) was carried out by \cite{Donnert2014MNRAS438}.
The parameters of the initial setup were based on those derived by \cite{MenanteauET2012ApJ748}.
\cite{Donnert2014MNRAS438} reproduced the X-ray luminosity of El Gordo and the observed distances between the peaks of the dark matter surface 
density distribution and X-ray emission and SZ effect. 
However, the X-ray morphology has not been reproduced well: the leading X-ray peak followed only by one tail close to the collision axis (the line connecting the two dark matter centers).
Donnert is suggesting that the reason why the two-tailed X-ray morphology was not reproduced was the lack of substructure in the simulation.

\smallskip
\section{FLASH Simulations for El Gordo}
\label{S:FLASH}

We carried out self-consistent numerical simulations of binary galaxy cluster mergers including dark matter and intracluster gas using the publicly available Eulerian parallel code, \FLASH, developed at the Center for Astrophysical Thermonuclear Flashes at the University of Chicago (\citealt{Fryxell2000ApJS131p273} and \citealt{Ricker2008ApJS176}). 
\FLASH, with its new modules, offers the possibility for the future to include many non-gravitational processes, such as
thermal conduction, viscosity, radiative cooling, and magnetohydrodynamics, which 
might be relevant for certain aspects of cluster merging.
The box size of our simulations was 13.3 Mpc on a side, reaching the highest resolution, i.e., cell size, of 12.7 kpc at the merger shocks and the centers of the clusters. 
Our simulations were semi--adiabatic, i.e., only shock heating was included.
In order to establish our notation and for the convenience of the reader, we briefly summarize our well established simulation method here,
a detailed description and verifications of the method can be found in \cite{Molnar2013ApJ779,Molnar2013ApJ774,Molnet2012ApJ748}.

\subsection{Initial setup}
\label{SS:INIT}

As initial conditions for the two colliding clusters, for the distribution of the dark matter and the intracluster gas, we assumed spherical models with a cut off at the the virial radius of each cluster ($r \le R_{\rm vir}$). 
We adopted a NFW model \citep{NFW1997ApJ490p493} for the dark matter distribution:
\begin{equation} \label{E:NFW}
      \rho_{DM} (r) =  { \rho_s  \over x (1 + x)^2}
,
\end{equation}
\nop
where $x = r/r_s$, and $\rho_s$, $r_s = R_{vir}/c_{vir}$ are scaling parameters for the density and radius, and $c_{vir}$ is the concentration parameter.
We assumed a truncated non-isothermal $\beta$ model for the gas density distribution,

\begin{equation}  \label{E:NFW}
      \rho (r) =  { \rho_0  \over (1 + y^2)^{3 \beta /2} }
,
\end{equation}
\nop 
where $y = r/r_{core}$, $\rho_0$, is the density at the center, and $r_{core}$ is the scale radius for the gas distribution. 
The temperature of the gas was determined from the equation of \HE\ assuming the ideal gas equation of state with $\gamma = 5/3$. 
We adopted $\beta=1$ for the large scale distribution of the intracluster gas (suggested by cosmological numerical simulations; e.g., \citealt{Molnet10ApJ723p1272}).

We represented the dark matter and the stellar matter in galaxies with particles since galaxies can also be considered collisionless for our purposes. 
The number of particles at each AMR cell was determined by the local density, normalized according the total number of particles, which were 5 million in our simulations, assuming a gas fraction of 0.14.

The velocities of the dark matter particles were derived from sampling a Maxwellian distribution with a velocity dispersion as a function of distance from the cluster center obtained from the Jeans equation \citep{LokasMamon2001MNRAS321} adopting an isotropic velocity dispersion.
The direction of the velocity vectors were assumed to be isotropic (for more details on modeling the particle velocity distribution see \citealt{Molnet2012ApJ748}).

\subsection{\FLASH\ Runs}
\label{SS:RUNS}

We performed a series of numerical simulations with a total mass of 
$2.15 \times 10^{15}$ ${\rm M}_{\odot}$ and a mass ratio of 1.9 
motivated by the previous analysis of El Gordo based on observations \citep{MenanteauET2012ApJ748}.
We adopted concentration parameters of $c_1 = 8$ and $c_2 = 9$, and core radii 
$r_{core1}$ = 288 kpc and $r_{core2}$ = 216 kpc, for the main and the infalling component
motivated by our previous merging cluster simulations producing offsets between peaks of 
dark matter surface density and X-ray emission \citep{Molnet2012ApJ748}.
We used different 
impact parameters, $P$, and infall velocities, $V_0$, searching for initial parameters 
which could reproduce the observed morphology of the X-ray, SZ and surface mass density of El Gordo.
The initial parameters for our simulations 
are listed in Table~\ref{T:TABLE1}. 
The IDs of our runs are listed in the first column as $RnPijVkl$, where
$n$ is the serial number of the run, the pairs of integers, $ij$ and $kl$, 
indicate the impact parameter, $P$ (in units of 10 kpc), and the infall velocity of the run, 
$V_0$ (truncated at the 100s in \KMSEC).
The second column lists the relative initial velocities of the infalling cluster in \KMSEC\ at the distance where the two clusters' truncated gas spheres touch. The third column lists the impact parameter, defined as the perpendicular distance between the trajectory of the infalling cluster and the center of the main cluster (in kpc).
The other columns show parameters related to the analysis of our simulations to be discussed
in Section~\ref{S:Results}.

%
%
\begin{deluxetable}{ccrcrrr}
\tablecolumns{7}
\tablecaption{
 \label{T:TABLE1} 
 Parameters
} 
\tablewidth{0pt} 
\tablehead{ 
 \multicolumn{1}{c}{ID\tablenotemark{a}} &
 \multicolumn{1}{c}{V$_0$\tablenotemark{b}} &
 \multicolumn{1}{c}{P\tablenotemark{c}} &
 \multicolumn{1}{c}{$\theta$\tablenotemark{d}} &
 \multicolumn{1}{c}{$\varphi$\tablenotemark{e}} &
 \multicolumn{1}{c}{D$_{p}$\tablenotemark{f}} &
 \multicolumn{1}{c}{V$_r$\tablenotemark{g}} 
 }
 \startdata  
    R1P05V22  &    2250   &       50$\;\;$      &   48\DEGREE  &    -60\DEGREE   &   860$\,$    &  655$\;\;\;\,$    \\ \hline
    R2P20V22  &    2250   &     200$\;\;$      &   49\DEGREE  & -160\DEGREE    &   850    &   690$\;\;\;\,$     \\ \hline
    R3P30V22  &    2250   &     300$\;\;$      &   45\DEGREE  &       5\DEGREE    &   914    &   712$\;\;\;\,$     \\ \hline
    R4P35V22  &    2250   &     350$\;\;$      &   44\DEGREE  &     10\DEGREE    &   932    &   745$\;\;\;\,$     \\ \hline
    R5P25V20  &    2000    &    250$\;\;$      &   50\DEGREE  & -160\DEGREE     &   760   &   497$\;\;\;\,$    \\ \hline
    R6P25V25  &    2500    &    250$\;\;$      &   44\DEGREE  &     10\DEGREE     &   960   &    935$\;\;\;\,$    \\ \hline
    R7P25V30  &    3000    &    250$\;\;$      &   42\DEGREE  &        8\DEGREE    &   960   &  1376$\;\;\;\,$    \\ \hline
    R8P10V25  &    2500   &     100$\;\;$      &   48\DEGREE  &    -60\DEGREE    &   860    &  1000$\;\;\;\,$     \\ \hline
    R9P20V25  &    2500    &    200$\;\;$      &   42\DEGREE  &      10\DEGREE    &   960   &    951$\;\;\;\,$ 
\enddata
\tablecomments{Parameters for initial conditions (columns 2 and 3), 
and for snapshots shown in our figures (columns 4-7).}
\tablenotetext{a}{Cluster ID.}
\tablenotetext{b}{Infall velocity in \KMSEC.}
\tablenotetext{c}{Impact parameter in kpc.}
\tablenotetext{d}{Polar angle of projection.}
\tablenotetext{e}{Roll angle of the projection.}
\tablenotetext{f}{Projected distance between the two dark matter peaks in kpc.}
\tablenotetext{g}{Radial velocity in \KMSEC.}
\end{deluxetable}  

\smallskip
\subsection{Image Simulations}
\label{SS:IMAGES}

After each simulation, we generated images of the X--ray surface brightness, 
SZ amplitude, and total mass surface density distribution for a range 
of viewing angles at different phases of the collision (i.e. different output times).
Our starting point for each snapshot was a coordinate
system with the $z$ coordinate axis aligned with the two dark matter centers 
pointing towards the infalling cluster, and the main plane of the collision 
(the plane determined by the two cluster centers and the relative velocity vector
parallel to the $z$ axis) aligned with the $x$--$z$ plane, the $x$ axis pointing 
towards the offset of the infall velocity from the main cluster center.
First, we rotated the cluster around the $z$ axis with a roll angle $\varphi$, 
which we changed freely, then around the $x$ axis with 
a polar angle, $\theta$, determined by the constraint of the 
observed projected distance between the two mass centers 
(active rotation: we rotated the cluster not the coordinate system).
At the end, we project the cluster along the $y$ axis, assuming that the 
$y$ axis is parallel to the line of sight (LOS). 
In this arrangement the zero polar and roll angles 
($\theta = 0$; $\varphi = 0$) mean 
that we assume that the collision happened in the plane of the sky.

%
%
\begin{figure}[t]
\includegraphics[width=.46\textwidth]{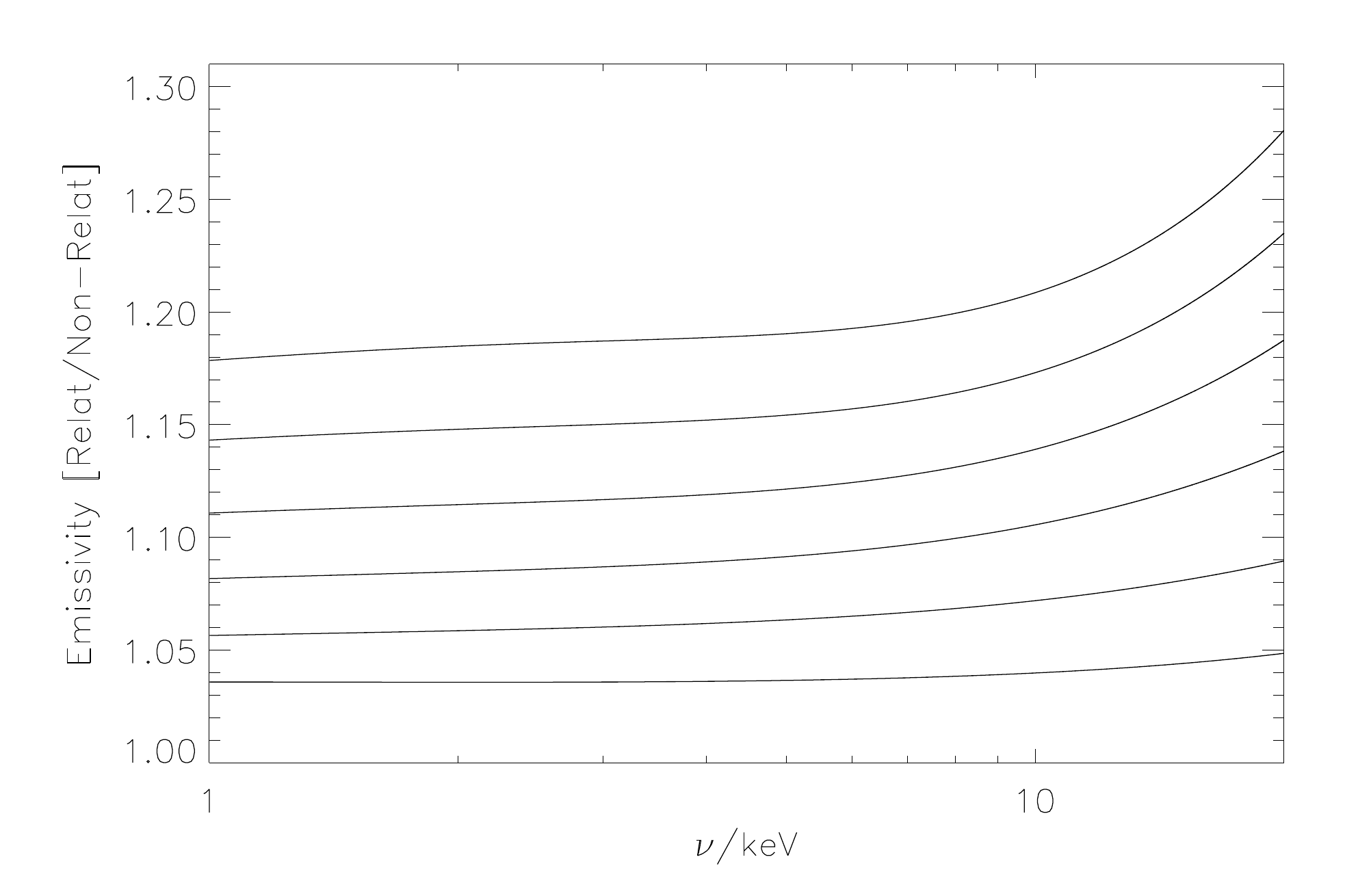}
\caption{
 Emissivity ratios: relativistic over non-relativistic as a function of frequency (in units of keV).
 The lines represent emissivities at different gas temperatures: 
 10, 20, 30, 40, 60, 80 keV (form bottom to top).
\label{F:RELATBREM}
}
\end{figure} 

Once we choose the time for the output and the rotation angles, we generated
X--ray surface brightness images by performing a LOS 
integral (along the $y$ axis) over the emissivity at fixed $x$ and $z$ using
\begin{equation}  \label{E:XRAY}
    I_X (x,z) \propto  \int  \, d y  \, \rho_g^2 \, \int   d \nu \,\varepsilon (Z_{ab}, T, \nu) \; A_{eff} (\nu)
,
\end{equation}
where $\varepsilon (Z_{ab},T,\nu)$ is the X-ray emissivity at frequency $\nu$, 
$Z_{ab}$ is the abundance, $T$ is the temperature of the intracluster gas, and 
$A_{eff}$ is the effective area of the telescope. 
We used the 4-chip averaged on axis effective area of ACIS-I \citep{ZhaoET2004}
as a sufficient approximation for our purposes for comparison with the \CHANDRA\ 
observations.

%
%
\begin{figure*}[t]
\includegraphics[width=1.00\textwidth]{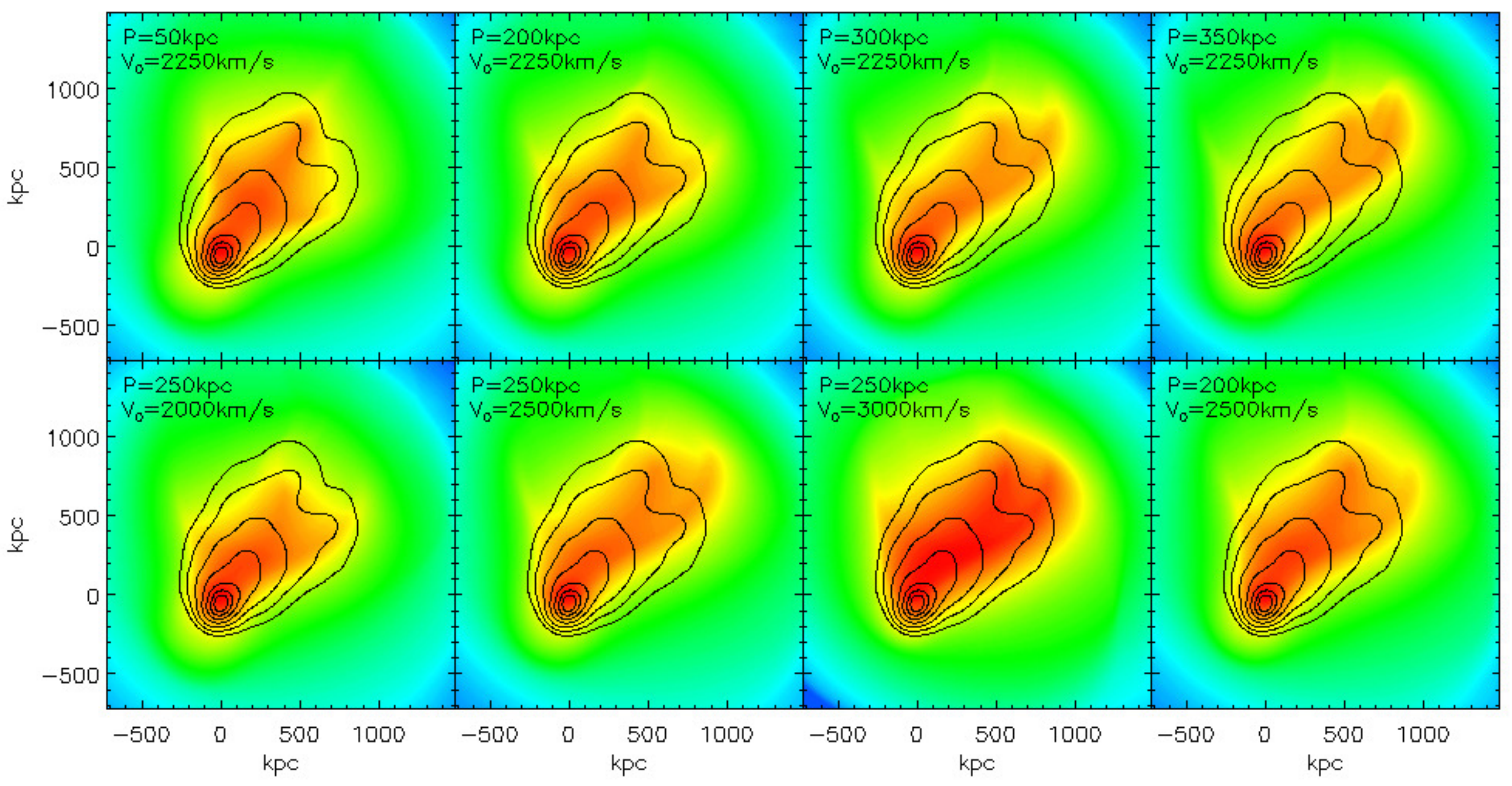}
\caption{
Examples of simulated X-ray surface brightness images from our merging cluster simulations 
with projection angles producing a cometary X-ray 
morphology and a wake with two tails behind the infalling cluster. 
In the upper row we show images from simulations with fixed infall velocity of 
$V_0=2250$ \KMSEC\ and different impact parameters
(runs R1P05V22, R2P20V22, R3P30V22, and R4P35V22; left to right).
In the lower row we show images from simulations with fixed impact parameters of 250 kpc and different
impact velocities, $V_0=2000$, 2500, 3000 \KMSEC, and a simulation with an impact parameter of 200 kpc 
and $V_0$ = 2500 \KMSEC\ (runs R5P25V20, R6P25V25, R7P25V30 and R9P20V25; left to right).
(See Table~\ref{T:TABLE1} for other parameters.) 
 Contours from \CHANDRA\ observations of El Gordo are overlaid for easier comparison 
 (courtesy of J. P. Hughes).
\label{F:XRIMAGES}
}
\end{figure*} 

X-ray emission from El Gordo was detected in the 0.3--9 keV frequency 
band (see Figure 3 of \citealt{MenanteauET2012ApJ748}), 
which corresponds to 0.561--16.8 keV at emission. 
The compressed, shock heated gas due to merging can reach 40--80 keV at
large impact velocities. At frequencies \simgreat 10 keV the relativistic
corrections increase the emissivity by 15--25\%, 
as opposed to only a few percent increase at a gas temperature of 10 keV;
overall, in the 0.5--17 keV frequency band, the emissivity of an 80 keV temperature 
gas is about 20\% higher than a 10 keV gas, as it can be seen from  
Figure~\ref{F:RELATBREM}. In this figure we show the emissivity ratio of
the relativistic to non-relativistic bremsstrahlung emissivity for gas temperatures of
10, 20, 30, 40, 60 and 80 keV as a function of frequency (in keV).
Even though we are not interested in the normalization of the X-ray emission,
relativistic corrections have to be included, because they change the relative
normalization: they enhance the emission for high temperature gas.
This enhancement is due to an increase in the relativistic Gaunt factor and the contribution from
electron--electron bremsstrahlung
(for a detailed discussion of the soft and hard X-ray emission due to thermal bremsstrahlung in 
merging galaxy clusters see Molnar et al. (2014, in preparation).
However, the commonly used X-ray spectral analysis packages 
(e.g., Xspec\footnote{http://heasarc.gsfc.nasa.gov}) have been developed for line diagnostics. 
The highest resolution X-ray emission package, APEC\footnote{http://www.atomdb.org}
includes some corrections, but, 
since it has been developed for line diagnostics, does not include all corrections.
Therefore we choose to calculate the continuum, $\varepsilon^C(\nu)$, and line X-ray emissivity,
$\varepsilon^L(\nu)$, at frequency $\nu$ separately: 

\begin{equation}
   \varepsilon (Z_{ab},T,\nu) \, d \nu  = \varepsilon^C(Z_{ab},T,\nu) + \varepsilon^L(Z_{ab},T,\nu) \, d \nu
,
\end{equation}
where, for the abundance of elements, $Z_{ab}$, we used 30\% of the Solar abundance
($Z_{ab} = 0.3 Z_\odot$).
For the continuum emissivity, we used the relativistic calculations of 
\cite{NozawaET1998ApJ507}: 

\begin{equation}
   \varepsilon^C (Z_{ab},T,\nu)\, d \nu = 
              \sum_{i=1}^N  \bigl[   \varepsilon_{ei}(n_e,n_i, T, \nu) + \varepsilon_{ee}(n_e,T,\nu) \bigr] \, d \nu
,
\end{equation}
where $n_e$ and $n_i$ are the electron and ion number densities, 
and the first term on the right hand side is a sum of emissivities over ions for electron--ion 
collisions, the second term represents the emissivity due to electron--electron collisions.
The first term can be expressed as 

\begin{equation}
    \varepsilon_{ei} \, d \nu  =  c_{ep} \, n_{e} n_i Z_i^{2} \,  g (Z_i, \Theta) \Theta^{-1/2} \, e^{-u} \, d \nu
, 
\end{equation}
where $c_{ep} = 16 \, ( 2 \pi/3 )^{3/2} \, m_e c^2 \, r_e^3$, $Z_i$, is the charge of ion $i$, 
the dimensionless temperature and frequency are 
$\Theta = k_B T / m_e \,c^2$, and $u = h_P \, \nu / k_B T$, where $k_B$ and $h_P$ are the Boltzmann and 
Planck constants, $m_e$ is the mass of the electron, $c$ is the speed of light,
and the thermally averaged Gaunt factor is calculated as
\begin{equation}
g (Z_i, \Theta) = \frac{\pi}{8} \left( \frac{3 }{2 \pi} \right)^{3/2} \Theta^{7/2} \, e^{u} \, 
                                        \frac{ J^{\,-}(\Theta, u, Z_{j})}{G_{0}^{\,-}(\Theta)}
,
\end{equation}
where $r_e$ is the classical electron radius, $r_e = e^2/ (m_e c^2)$,
(where $e$ is the charge of the electron), 
and the lengthy expressions for the $J^{\,-}(\Theta, u, Z_{j})$ 
and $G_{0}^{\,-}(\Theta)$ integrals, which are easily done numerically, 
can be found in \cite{NozawaET1998ApJ507}.

We used the fitting formula of \cite{Nozawa2009AA499} for the relativistic electron--electron 
thermal bremsstrahlung emissivity, 
\begin{equation}
   \varepsilon_{ee} \, d \nu  = 
              c_{ee} \,n_e^2\,\Theta^{-1/2} \, e^{-u} \, G^{II}_{PW}(\Theta, u)\, F^{II}_{CC}(\Theta, u)\,d \nu
, 
\end{equation}
where  $c_{ee} = \alpha \, \sigma_T \, m_e c^3$, where $\alpha$ is the fine structure constant, 
$\sigma_T$ is the Thomson cross section, and the fitting functions $G^{II}_{PW}(\Theta,u)$ and 
$F^{II}_{CC}(\Theta,u)$ can be found in \cite{Nozawa2009AA499}.
For the line emissivity, $\varepsilon^L(\nu)$, we used the 
publicly available APEC code developed for line diagnostics\footnote{http://www.atomdb.org}.

%
%
\begin{figure*}[t]
\includegraphics[width=1.00\textwidth]{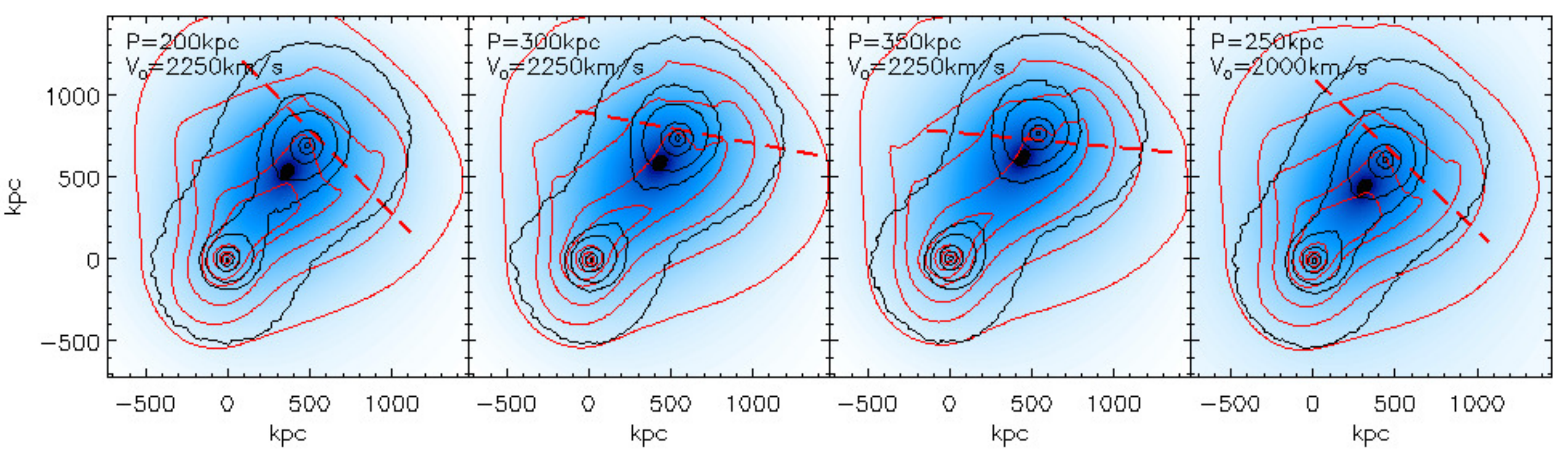}
\caption{
  X-ray surface brightness, SZ amplitude and total mass surface density contours (red, blue, and black)
  from our simulations with infall velocity of $V_0=2250$ \KMSEC\ and different 
  impact parameters: $P$~=~200, 300, 350 kpc (runs R2P20V22, R3P30V22, and R4P35V22), 
  and from a simulation with $V_0$~=~2000 \KMSEC\ and $P$ =~250 kpc
  (run R5P25V20; left to right). Table~\ref{T:TABLE1} contains information of other parameters.
  The normalizations and contour levels are arbitrarily chosen for easy comparison with observations.
  The X-ray surface brightness profiles for paths shown as red dashed lines are displayed in 
  Figure~\ref{F:XRAYPROFILES}.
\label{F:XRAYSZDMCONT}
}
\end{figure*} 

The SZ surface brightness was calculated using relativistic corrections, 
following \cite{itohet98ApJ502},

\begin{equation}  \label{E:XRAY}
    I_{SZ}\, (x,z) \propto  \int d y \,  \rho_g \, T_g
                                                     \biggl[ g(\nu) +  \sum_{n=1}^{n=4} Y_n \, \Theta_e^{\,n} \biggr]
,
\end{equation}
where $g(\nu) = \coth ( x_\nu/2) - 4$ is the non-relativistic frequency function,  
where the the dimensionless frequency is 
$x_\nu = h_P\, \nu / ( k_B T_{cmb})$, $T_{cmb}$ is 
the temperature of the cosmic microwave background, and the second term is a sum of 
the relativistic corrections approximated as a polynomial with coefficients, 
$Y_n$, listed in \cite{itohet98ApJ502}.

We integrated the total (dark matter and gas) density along the $y$ axis
to generate the mass surface density images at position, $x$ and $z$:

\begin{equation}  \label{E:SIGMA}
    \Sigma (x,z) \propto  \int d y \,   \bigl( \rho_d + \rho_g  \bigr)
,
\end{equation}
Note that the total mass surface density maps are dominated by the main mass component, the dark matter.

\smallskip
\section{Results}
\label{S:Results}

When carrying out numerical simulations of binary cluster collisions,
the observed mass surface density distribution provides a starting estimate for the 
initial masses of the two components. 
Although here we fix the masses to reduce the initial parameter space, they can also be determined along with other input parameters by comparing observations and numerical simulations.
The observed projected distance between the two mass centers, by itself, does not 
provide constraints on the phase of the collision 
(e.g., the time relative to the first core passage) due to possible projection effects.
The phase of the collision can be constrained by using multi-frequency observations
comparing the morphology of the intracluster gas derived from high resolution X-ray and/or SZ observations,
and the positions of the mass centers (in projection) derived from optical/infrared observations 
using the method of gravitational lensing (e.g., \citealt{Molnet2012ApJ748}).
In the case of El Gordo, the SZ observations have a low angular resolution of about 1.\AMIN4 (ACT).
Therefore, in order to constrain the morphology of the gas, we can only use
the \CHANDRA\ X-ray observations which have a resolution of $\sim$1\ASEC.

\subsection{Reproducing the two-tail morphology}

In \cite{Molnet2012ApJ748} we performed a set of simulations to study the offsets between the X-ray emission, the SZ amplitude and mass surface density peaks. Based on a comparison with that study, we conclude that El Gordo is most likely after the first core passage, and before the turn around aiming for the second core passage: 
Before the first core passage there should have been two distinct, smooth X-ray peaks where the gas is in equilibrium around them. It must be that the core has passed for there to be only one X-ray peak and the gas has a very disturbed, cometary morphology.
Therefore, in order to reduce the parameter space, 
we fixed the initial masses of the two components as 
M$_1 = 1.4 \times 10^{15}$ ${\rm M}_{\odot}$ and 
M$_2 =  7.5 \times 10^{14}$ ${\rm M}_{\odot}$ (mass ratio of 1.9), 
and ran a set of simulations changing the other parameters looking for a strongly 
peaked X-ray emission with a cometary structure, and two tails 
in the wake behind the collision after the first core passage.

We found that a series of simulations with 
different infall velocities $\simgreat\, 2000$ \KMSEC\ 
and in a relatively narrow range of impact parameters, about 200 to 350 kpc, 
were able to reproduce the X-ray morphology with the two tails in the wake.
Simulations with small impact parameters produce three X-ray tails
(see Section~\ref{S:Discussion}), 
very large impact parameters result in only one tail.
In Figure~\ref{F:XRIMAGES}, as an illustration, we show
examples of simulated X-ray surface brightness images from our merger simulations, displayed with  
projection angles that match best the cometary X-ray 
morphology with a "twin tailed''  morphology behind the infalling cluster
as well as its physical size ($\sim$860 kpc), 
except the first panel in the upper row which is included for our discussion presented in Section~\ref{S:Discussion}
(here we ignore very faint additional tails).
In this Figure we overlaid the contours from \CHANDRA\ observations of El Gordo 
\citep{MenanteauET2012ApJ748} for an easy comparison.
In the upper row we show images from simulations with fixed infall velocity of $V_0$~=~2250 \KMSEC, 
but different impact parameters, 
$P=50$, 200, 300, 350 kpc (runs R1P05V22, R2P20V22, R3P30V22, and R4P35V22; left to right).
In the lower row we instead fix the impact parameter to 250 kpc and vary the
infall velocities, $V_0=2000$, 2500, 3000 \KMSEC\ 
(runs R5P25V20, R6P25V25, and R7P25V30; first 3 panels, left to right).
The last panel in the lower row is from a simulation with $P$ = 200 kpc and $V_0$ = 2500 \KMSEC\ 
(run R9P20V25). 
The IDs for the runs, the infall velocities, $V_0$, the impact parameters, $P$, and the 
polar and roll angles, $\theta$ and $\varphi$, of the corresponding rotations are shown in 
Table~\ref{T:TABLE1}.
The other two parameters shown in this table, the projected distance
between the two dark matter centers and the radial velocity associated with each projection
(columns 6 and 7), are discussed in Section~\ref{SS:VRAD}.

We find that the phases of the collision for these massive clusters with large
infall velocities (\simgreat \,2000 \KMSEC) and small mass ratios are the following:
First a large scale contact discontinuity forms when 
the gas of the infalling cluster presses against that of the main cluster. 
As the infalling cluster plunges into the main cluster,
adiabatically compressed layers of gas develop in both components around the contact 
discontinuity, and form a wedge-shaped region, 
and a shock develops on the two sides of the contact discontinuity.
This wedge-shaped contact discontinuity passes through the main cluster 
and visible after the first core passage as well (upper row, first panel in Figure~\ref{F:XRIMAGES}).
However, in the case of El Gordo, due to an impact parameter in the order of a few hundred kpc,
and a high infall velocity, only one side of the wedge-shaped region is visible in the X-ray image, 
because there is not enough gas on the side of the collision to produce 
a visible signal (e.g., first row 3rd, 4th panel in Figure~\ref{F:XRIMAGES}).
After the first core passage, a bow shock forms in the main cluster gas 
ahead of the contact discontinuity as the infalling cluster makes its way 
outward from the main cluster (clearly visible in all panels of Figure~\ref{F:XRIMAGES}).
This is the same phase as that of the Bullet cluster.
At this stage of the collision, tidally stretched gas produces the X-ray emission that connects the two mass centers, 
seen 
as the Northern X-ray tail in the wake of El Gordo.
This kind of tail can always be seen in simulations with smaller infall velocities, 
because in this case the gravitational force of the main cluster 
can retain enough gas to produce visible X-ray emission. 
In our simulated X-ray images there is a hint of a third tail, above the tidally stretched gas, 
which is due to the compressed gas on that side of the collision axis.
The X-ray emission observed in El Gordo suggests the presence of a possible third tail, indicated by enhanced emission to the North-West of the X-ray peak. However, the photon counts are quite low in this direction. 
Our simulations suggest that a third X-ray tail above the two observed tails would be
visible in an X-ray image of El Gordo with a longer exposure time.

\subsection{Reproducing projected distance and relative radial velocity}
\label{SS:VRAD}

We have demonstrated that the observed cometary X-ray morphology and system size of El Gordo can be reproduced by our simulations using several infall velocities and impact parameters values (Figure~\ref{F:XRIMAGES}). 
As a next step, we use the observed projected distance between the two dark matter 
centers and the observed relative radial velocity of the infalling cluster to constrain the 
initial parameters of the system. 
The mass center of the main (NW) component marked by 
the rest frame $i$-band and 3.6~$\mu$m luminosity density, stellar mass density
\citep{MenanteauET2012ApJ748}, and mass surface density derived from gravitational lensing 
\citep{JeeET2013arXiv1309.5097} are all very close to each other, so they seem to be reliable.
On the other hand, the position of the mass peak of the infalling cluster, the SE component, 
derived using different methods show about a 170 kpc scatter around the X-ray peak.
Since our simulations suggest that the X-ray peak and the mass peak of the infalling
cluster should be close, we adopt the position of the X-ray peak as the center
of the infalling cluster. Thus we demand that the projected distance 
between the centers of the two components in our merging simulations to be $\sim$860 kpc. 
Based on galaxy redshift measurements, the observed relative radial
(LOS) velocity between the infalling cluster center and
the main component was found to be 586$\pm$96 \KMSEC\  \citep{MenanteauET2012ApJ748}.
However, the observed radial velocity of the BCG was 731$\pm$66 \KMSEC, 
so we may assume that the radial relative velocity in El Gordo falls in the interval of 
500--800 \KMSEC.

The projected distances between the mass centers and the relative radial
velocities for our simulations are listed in Table~\ref{T:TABLE1}
(columns 6 and 7).
Based on this table, we conclude, that simulations with an infall velocity of
$V_0=2250$ \KMSEC\ and impact parameters $P=200$, 300, and 350 kpc 
(runs R2P20V22, R3P30V22, and R4P35V22), and $V_0~=~2000$ \KMSEC\ and $P~=~250$ kpc
(run R5P25V20) provide a good match with the X-ray morphology, the positions of the 
X-ray peak and mass peaks, and the constraints from the observed radial relative velocity. 
We can exclude run R1P05V22, because it has three tails due to the small impact parameter 
of $P~=~50$ kpc (we included only for comparison with the simulation carried out by
\citealt{Donnert2014MNRAS438} for El Gordo).
In general we can exclude simulations with small
impact parameters because they produce either one or three tails depending on the infall velocities.
Runs R6P25V25, R7P25V30, and R9P20V25 can be excluded as well, because these simulations 
with larger infall velocities produce either too large projected 
distances between the two mass centers, or too large radial velocities.
In Figure~\ref{F:XRAYSZDMCONT} we show the X-ray surface brightness,
SZ and mass surface density contours for these simulations 
(runs R2P20V22, R3P30V22, R4P35V22, and R5P25V20) in the same 
projection as in Figure~\ref{F:XRIMAGES} (see Table~\ref{T:TABLE1} for the parameters). 
All four contour images show similar morphology. 
The X-ray and mass contours show that the mass centers of the ``bullet'' coincide with the X-ray peak. At these assumed infall velocities, unlike in the case of the Bullet cluster, the ram pressure is not sufficient to drag the gas behind the dark matter component as it passes through the main cluster gas cloud.
Thus our simulations predict that we will not find an offset between the X-ray peak 
and the mass center of the ``bullet'' using more precise observations. 
On each image, the mass center of the main cluster is located at the North-Western 
end of the Northern tail, as predicted by observations.

\subsection{Reproducing the X-ray brightness profile}

\begin{figure}[t]
\includegraphics[width=.46\textwidth]{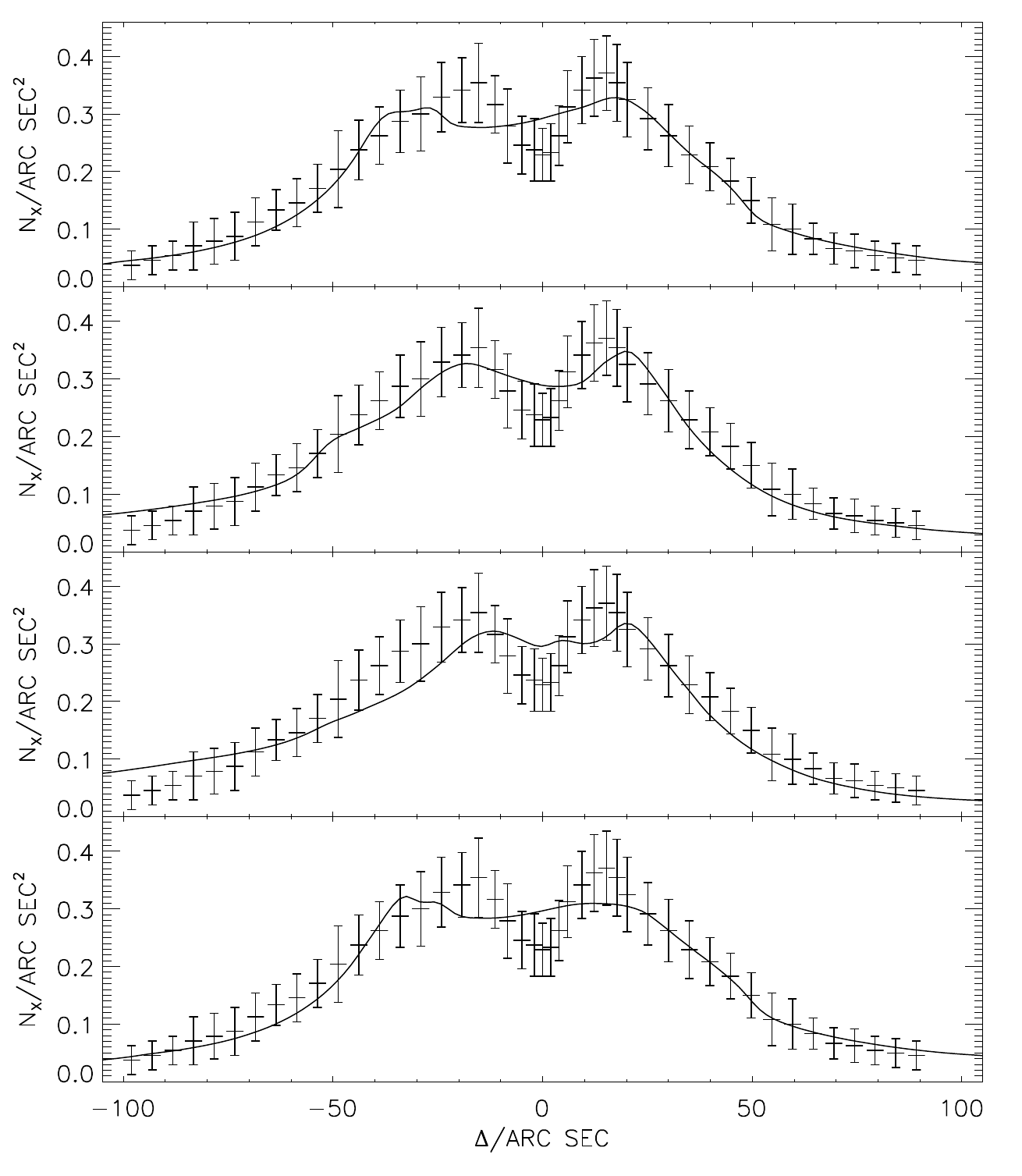}
\caption{
    Time integrated X-ray surface brightness (in photon counts, $N_X$, per arc sec$^2$)
    across the wake of the cluster merger taken from the Chandra 
    observations (data points with error bars) overlaid on those extracted from our FLASH simulation (solid lines). 
    The simulations were done with an infall velocity of $V_0$~=~2250 \KMSEC\ and different impact parameters, 
    $P$~=~200, 300, 350 kpc (runs R2P20V22, R3P30V22, and R4P35V22) and with
    $V_0$~=~2000 \KMSEC\ and $P$ = 250 kpc (run R5P25V20) 
    as a function of distance along the path, $\Delta$, in arc second (top to bottom).
    The same projections of these simulations are shown in Figure~\ref{F:XRAYSZDMCONT}.
    The extraction path for these profiles (shown as red dashed lines in Figure~\ref{F:XRAYSZDMCONT})
    were chosen to provide the best match with the X-ray surface brightness profile across the wake of El Gordo 
    derived from \CHANDRA\ observations.
\label{F:XRAYPROFILES}
}
\end{figure}

We further constrain the infall velocity and the impact parameter of the collision for
El Gordo comparing the shapes of our X-ray surface brightness profiles across the wake 
with that from \CHANDRA\ observations (Figure 1 of \citealt{MenanteauET2012ApJ748}).
The observed profile is nearly symmetric with two maxima separated by about 35\ASEC,
and a local minimum between them. 
First, we went through different snapshots for each run after the first core passage with a separation
between the two dark matter centers of 600--1200 kpc, and generated X-ray images by projecting 
them out with a set of different polar and roll angles on a grid, and kept those snapshots and 
projections which resembled El Gordo (showed a cometary morphology with one peak X-ray emission 
and two tails in the wake of the infalling cluster). As a next step, we extracted X-ray surface brightness 
profiles along cuts through the wake with different positions and rotation angles, and found the best
fits to the \CHANDRA\ data. The extracted X-ray surface brightness profiles of those 
projections of each run that produced the best match with El Gordo observations are shown in 
Figure~\ref{F:XRAYPROFILES} (runs R2P20V22, R3P30V22, R4P35V22, and R5P25V20). 
The X-ray images for these snapshots with the same projections 
are shown in Figure~\ref{F:XRIMAGES} (upper row panels 2, 3 and 4, and lower row first panel).
The corresponding paths are displayed in Figure~\ref{F:XRAYSZDMCONT} (red dashed lines).
In Figure~\ref{F:XRAYPROFILES} the El Gordo observations are also shown 
with points with error bars (read off from Figure 1 of \citealt{MenanteauET2012ApJ748}).
Our simulation with $V_0=2250$ \KMSEC, and $P=300$ kpc
is the best match with the observed profile. Our simulation with 
$V_0=2250$ \KMSEC\ and a smaller impact parameter, $P=200$ kpc, 
and a simulation with $V_0=2000$ \KMSEC\ and $P=250$ kpc
(runs R2P20V22 and R5P25V20) have the two peaks too far from each other, 
the profile from our simulation with $V_0=2250$ \KMSEC\ and larger impact parameter, 
$P=350$ \KMSEC\ (run R4P35V22) has three X-ray peaks in the wake.
Therefore, based on the observed X-ray morphology, the profile of the X-ray surface density
across the wake, the relative radial velocity, and the relative positions
of the X-ray, SZ and mass peaks, subject to the limitations of our models, we derive an
infall velocity of $V_0=2250\pm250$ \KMSEC, and 
a large impact parameter of $P=300_{-100}^{+50}$ kpc for El Gordo
(note, the errors shown are rough estimates only).

%
%

\section{Discussion}
\label{S:Discussion}

Our results confirm the basic interpretation of the merging cluster El Gordo 
by \cite{MenanteauET2012ApJ748}: this system is after the first core passage
of a large infall velocity collision, the infalling cluster moving from the 
NW towards SE, and the main cluster center coincides with the observed position of the 
mass concentration of the NW component.
Based on the merger morphology, Menanteau et al. (2012) proposed that the inclination 
angle (between the collision axis and the plane of the sky) should be shallow.
Assuming that the infall velocity is close to the velocity of the observation after the
first core passage, Menanteau et al. predicted an infall velocity of 2300 \KMSEC\ and 1200 \KMSEC\
for polar angles of 15\DEGREE\ and 30\DEGREE.
Our best result with an infall velocity of 2250 \KMSEC\ (run R4P35V22) is in 
a good agreement with their prediction, however this is a coincidence.
The relative velocity of this run at the phase of best fit with observations is 1009 \KMSEC. 
This shows that the instantaneous velocity at a phase
after the first core passage could be 50\% of the infall velocity
(or it could even be zero at the phase of the turnaround). 
The rotation angle out of the sky (the main plane of the collision), i.e., the polar angle, $\theta$,
in our case is 45\DEGREE. Following Menanteau et al.'s procedure, 
using this rotation angle, we would find 600/$\sin 45$\DEGREE $\approx\,$850 \KMSEC\
for the infall velocity, and not 2250 \KMSEC.

%
%
\begin{figure}[t]
\includegraphics[width=.46\textwidth]{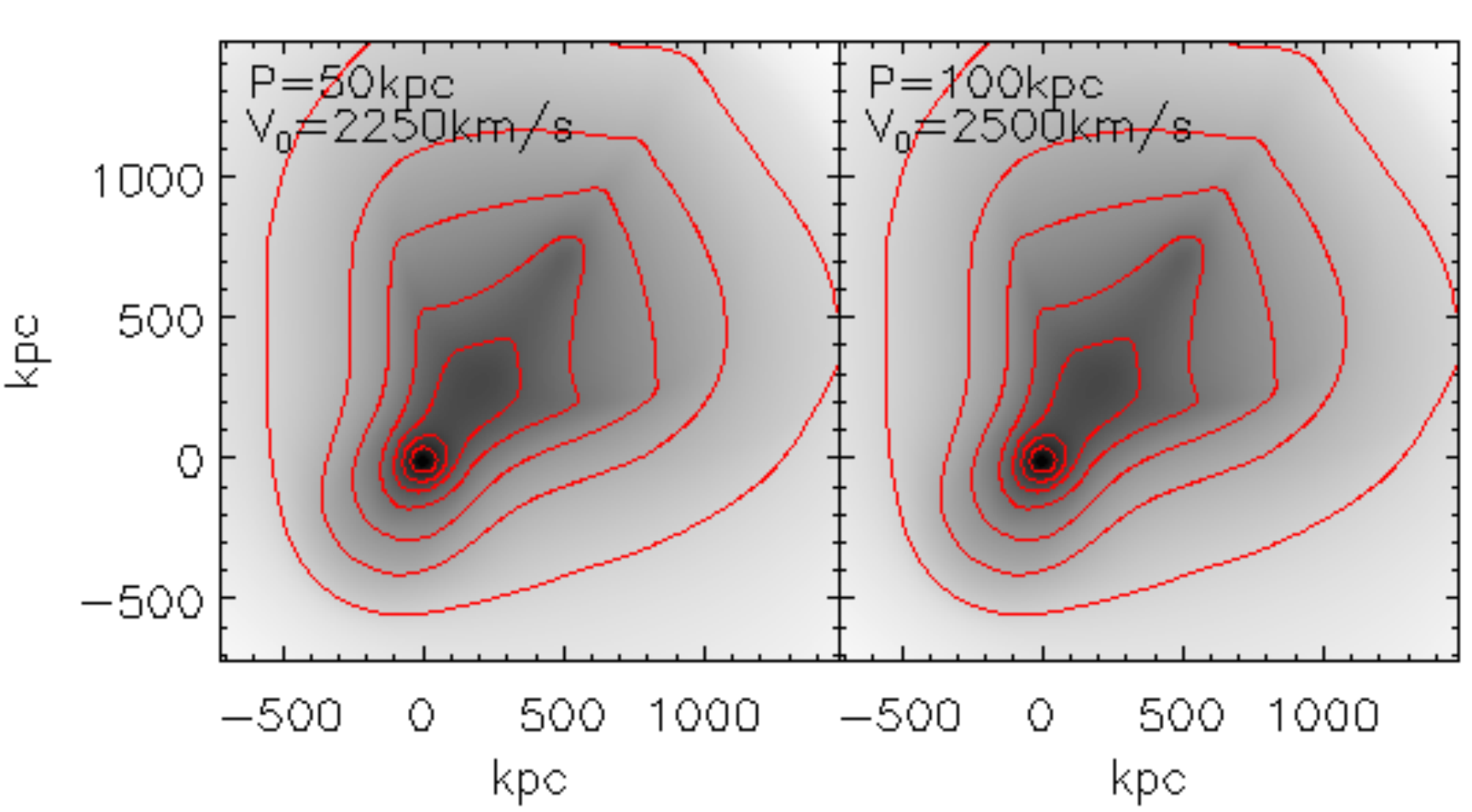}
\caption{
    X-ray surface brightness images from our \FLASH\ simulations 
    with small impact parameters, $P=50$ and 100 kpc, and infall velocities of $V_0=2250$ and 2500 \KMSEC\ 
    (runs R1P05V22 and R8P10V25; left to right). For other parameters see Table~\ref{T:TABLE1}.  
    The contours of the same data are overlaid to emphasize the morphology.
\label{F:XRAYCOMP}
}
\end{figure} 

\cite{Donnert2014MNRAS438} carried out simulations to reproduce 
the observed properties of El Gordo. 
Most of the important initial conditions adopted were similar to ours 
(we also based our simulations on the results of \citealt{MenanteauET2012ApJ748}):
large masses and the core radius for the main component:
M$_1 = 1.9 \times 10^{15}$ ${\rm M}_{\odot}$, 
M$_2 =  8.1 \times 10^{14}$ ${\rm M}_{\odot}$, 
$r_{core1}$ = 300 kpc, infall velocity: 2600 \KMSEC. However, 
the impact parameter, the concentration parameters, $\beta$, 
and the core radius of the infalling cluster were significantly smaller than ours: 
20 kpc, $c_1$~=~2.9, $c_2$~=~3.2, $\beta$ = 2/3, $r_{core2}$ = 25 kpc, as well as $\beta$~=~2/3, 
and the initial distance, which does not affect the results, larger: 5231~kpc.

Donnert used a Hernquist profile \citep{Hernquist1990ApJ356} for the distribution of dark matter 
without a cut off at the virial radius, but since the density profile of this model falls off as $r^{-4}$, 
the density is quite low at large radii, whereas we used a truncated NFW profile with 
zero density at large radii.
This difference should not have any significant effect on neither 
the dynamics of the collision nor the observable X-ray emission around the first core passage.
Donnert's simulations reproduced the X-ray luminosity of El Gordo, the distances between 
peaks of the X-ray emission, SZ signal and the mass surface density well. 
However, the X-ray morphology was not reproduced correctly: the cometary structure showed 
only one tail pointing to the center of the mass distribution of the main cluster. 
Our simulations were able to reproduce the two X-ray tails
with a range of infall velocities (2000--3000 \KMSEC) and impact parameters (200--350 kpc).

Turbulence plays an important role in structure formation, it facilitates the mixing of the intracluster gas during cluster merging.
Comparisons between SPH and Eulerian AMR codes suggest that artificial viscosity, necessary in SPH codes to handle shocks, suppress dynamical instabilities, and thus turbulent mixing of the gas (Mitchell et al. 2009; Agertz et al. 2007). 
Therefore, as opposed to Donnert, we chose to use an AMR code for our merging cluster simulations. 
However, the main physical process which produces ordered large scale shock-heated and compressed gas
is more likely linked to formation of contact discontinuities and not to turbulence.

In Figure~\ref{F:XRAYCOMP} we show X-ray images from our merging cluster simulations 
with an infall velocity of $V_0=2250$ \KMSEC\ and an impact parameter of $P=50$ kpc, 
and with $V_0=2500$ \KMSEC\ and $P=100$ kpc (runs R1P05V22 and R8P10V25; left to right).
We overlaid the contours of the same X-ray images to emphasize the morphology. 
These images show a similar symmetric morphology to that of Donnert, 
who used $V_0=2600$ \KMSEC\ and $P=20$ kpc.
We may conclude that even an impact parameter of 50 or 100 kpc 
would not break the symmetry of the X-ray images significantly, and can not produce
the asymmetric two-tail morphology of El Gordo.
In both of these panels of Figure~\ref{F:XRAYCOMP}, three tails can be seen, 
a larger tail in the middle and two fainter tails in a wedge shape pointing towards the bullet.
Our simulations suggest that the one tail seen in the wake of Donnert's simulation is due to the tidally 
stretched gas between the two mass centers. A similar wedge-shaped structure is generated in 
Donnert's simulation, but it disappears after the first core passage (Figure 5 of \citealt{Donnert2014MNRAS438}).
We speculate that Donnert's simulation with a small impact parameter 
produced only one X-ray tail and not three since they assumed the infalling cluster is in a bubble inside the main cluster atmosphere. This causes their contact discontinuity to form only between the two cores and not on a large scale of the entire cluster, as in our simulations 
\citep{Molnar2013ApJ779,Molnet2012ApJ748}.
In our simulations, we needed impact parameters of a few 100 kpc to produce two asymmetric X-ray tails.
We note that our simulation, R8P10V25, with the same infall velocity, 2500 \KMSEC, used by Donnert and with a small impact parameter, predicts a relative radial velocity of 1000 \KMSEC, much larger than observed (Table~\ref{T:TABLE1}).

\smallskip
\section{Conclusions}
\label{S:Conclusion}

Using self-consistent N-body/hydrodynamical simulations we confirm the previous 
interpretation of El Gordo, as a massive galaxy cluster in an early stage of merging, about
480 million years after a first core passage, well before the first turn around.
We have used the X-ray morphology, SZ, gravitational lensing and radio observations 
of El Gordo to constrain the initial conditions of this merging system. 
We have found that our simulations within a relatively narrow 
range of infall velocities and impact parameters can reproduce the data well.
In particular, we can reproduce the distinctive ``twin-tailed'' cometary X-ray morphology 
of El Gordo with a simple clear physical explanation. 
We match the observations with an initial infall velocity of $\simeq\,$ 2250 \KMSEC, 
when the clusters are separated by the sum of their virial radii. 
The impact parameter of the collision is well constrained at $\simeq\,$300 kpc
to provide the asymmetric second tail which is a projection of the compressed and shock heated gas front. 
We quantify this using the profile of the X-ray surface brightness across the wake
\citep{MenanteauET10}, by the relative radial velocity from spectroscopy of member galaxies, and
from the relative positions of the X-ray, SZ and mass peaks. 
Note that our best simulation does reconstruct all the observed morphological features of El Gordo, 
as well as the X-ray surface brightness across the wake quantitatively, 
but the position of the extraction region through the wake is not an exact match with that of 
the \CHANDRA\ observations. Therefore, for a more accurate determination of the 
infall velocity, we need to perform a search in a larger parameter space including 
the masses of the components, their concentration parameters, core radii for the intracluster gas.
This task necessitates many more simulations, which we defer to a future study.

This massive merging cluster with an infall velocity of 2250~\KMSEC\ at the the time the 
Universe was half of its age (at the redshift of $z=0.87$) is apparently very unusual in the 
standard \LCDM\ models, for which extensive simulations have been performed to determine
the expected probability distribution of relative velocities \citep{ThomNaga2012MNRAS419}.
At face value, such extreme cases present a challenge to our understanding of structure 
formation and may lead to a better understanding of dark matter and/or alternative theories of gravity. 
We aim to expand our grid of input parameters for a fuller investigation of the 
uniqueness of our solution and the uncertainties on the derived physical parameters using  
more powerful GPU accelerated architectures. 
We will also benefit from deeper X-ray observations and higher resolution SZ imaging to more accurately constrain our models. 
This work will become increasingly more important for cosmology as new discoveries of extreme cluster collisions are uncovered in the new wide area SZ surveys sensitive to high pressure shocked gas and also from future radio surveys sensitive to ``relic'' radio emission from merger shocks.

\acknowledgements
The code FLASH used in this work was in part developed by the
DOE-supported ASC/Alliance Center for Astrophysical Thermonuclear
Flashes at the University of Chicago.  We thank the Theoretical
Institute for Advanced Research in Astrophysics, Academia Sinica, for
allowing us to use its high performance computer facility for our simulations.  
We thank J. P. Hughes for providing us the processed 
X-ray data from their \CHANDRA\ observations of El Gordo, and 
the referee for a thorough reading of our paper and for
suggestions, which helped to improve on the presentation of our results.
This research has made use of the NASA/IPAC
Extragalactic Database (NED) which is operated by the Jet Propulsion
Laboratory, California Institute of Technology, under contract with
the National Aeronautics and Space Administration.

%
%
\bibliographystyle{apj}


\end{document}